\documentstyle[12pt]{article}

\begin{document}

\LaTeX{}\bigskip\ \bigskip\ 

\begin{center}
Dense matter theory: a simple classical approach\bigskip\ 

P.Savic$^1$ and V.Celebonovic$^2$\bigskip\ 

$^1$ Serbian Academy of Sciences and Arts,Knez Mihajlova 35,11000
Beograd,Yugoslavia

$^2$Institute of Physics,Pregrevica 118,11080 Zemun-Beograd,Yugoslavia.%
\medskip\ 

celebonovic@exp.phy.bg.ac.yu \smallskip\ 

vcelebonovic@sezampro.yu\bigskip\ \bigskip\ \bigskip\ \bigskip\ \bigskip\ 
\bigskip\ 

$^2$to whom all communications should be sent.
\end{center}

\newpage\ 

Abstract: In the sixties,the first author and R.Kasanin started developing a
mean-field theory of dense matter.It is based on the Coulomb
interaction,supplemented by a microscopic selection rule and a set of
experimentally founded postulates.Applications of the theory range from the
calculation of models of planetary internal structure to DAC experiments.

\newpage\ 

The purpose of this paper is to review briefly the main ideas and examples
of applicability of a particular classical theory of the behaviour of
materials under high pressure.It was developed jointly by the first author
and R.Kasanin in the early sixties [2] (abbreviated as SK in the following ).

The starting point for its development was a paper by Savic [1], referred to
as S61, which had the aim of exploring the origin of rotation of celestial
bodies.As a result of this work, there emerged the conclusion that rotation
is closely related to the internal structure,and that a theory of dense
matter was needed to explain it correctly.

The starting object of S61 is a low-temperature cloud of arbitrary shape
containing any number of chemical elements and their compounds.Two processes
influence the life of such a cloud: the mutual gravitational interaction of
its constituting particles,and the loss of energy due to thermal radiation.
As a combined result,the temperature of the cloud decreases,while its mean
density and internal pressure increase.Increasing pressure leads to
excitation and ionisation of atoms and moleules in its interior.In
quantum-mechanical terms,this means that increasing pressure leads to the
expansion of the radial part of the elctronic wave-function of the atoms and
molecules that make up the material. A quantum-mechanical treatment of this
process has been given only recently [3],nearly three decades after the idea
was used in SK .

Due to pressure ionization,the primeval cloud passes into the state of a
two-component plasma (assuming,for simplicity,that the cloud consists of one
chemical element ). It consists of a randomly moving free-electron gas (
which has a non-zero magnetic field [4]) and the atoms and molecules ionized
under pressure.Owing to high pressure their magnetic moments become oriented
in parallel ,and the resulting torque starts the rotation of the whole
system. Although it may seem highly qualitative,a detailed elaboration of
this mechanism ( [2] parts III and IV , or [5] ) gives values of the
strengths of the magnetic fields and the allowed intervals of the speed of
rotation of the Sun and the planets, which are in good agreement with the
observed values. For example,the SK theory gives for the magnetic field of
Jupiter a value between 10 and 14 Gauss,while the measured value is 14 [5]
.The observed value of the speed of rotation of the solar equatorial region
is 2.9 * 10$^{-6}$ rad s$^{-1}$ ; the possible interval according to SK is (
1.2 $\leq \omega \leq $ 44.7 ) rad s$^{-1}$.\newpage\ 

Apart from the magnetic fields and the speed of rotation, SK gives the
possibility of complete modelling of the internal structure of solar system
bodies. One can thus determine the number of layers in the interior of the
object and their thickness,the distribution of pressure,density and
temperature with depth,the mean atomic mass of the cehmical mixture that the
object is made of. For example,it was calculated within SK that the depth of
the Moho discontinuity is 39 km; the experimental value is 33 km. It was
shown that the magnetic moment of the Moon is zero, which was later
confirmed by in-situ measurements. More examples are given in [5] . The
temperature of the Earth's center was estimated,starting from SK , as 7000 K
[9] , which is close to experimental values. A word about chemistry: it has
been shown that the asteroid (1) Ceres and Neptune's satellite Triton are
similar ( by their mean atomic mass ) to Mars and Mercury, which has
cosmogonical implications .

The SK theory has also found applications in laboratory high pressure
work.It provides a method for determining phase transition points and
equations of state of materials exposed to high pressure by a calculational
procedure that is much simpler than the usual approach in statistical
mechanics.The mean interparticle distance is defined so as to correspond to
the position of stable equilibrium of the ''full'' interparticle
interaction; it is assumed in SK that the atoms and/or molecules in a
specimen under pressure interact only by the bare Coulomb potential. A
succession of phase transitions is presumed to occur in a specimen subdued
to increasing pressure,and a selection rule,giving the possibility to
''pick'' only those transitions which are physically realizable in a given
material, has been developed. The densities of two successive phases are
assumed to differ by a factor of two. This ratio is a consequence of an
empirical rule,first discussed in S61 in an astrophysical context whose
validity was later extended to laboratory high pressure work [2],[8],[10].

Starting from these ( and three more ) postulates discussed in detail in
[8],it becomes possible to determine high pressure phase transition points
of materials in DAC experiments .\newpage\ 

Metallisation of hydrogen is predicted to occur at 3 Mbar,while the
corresponding value for helium is 106 Mbar [8]. A detailed comparison of the
predictions of the SK theory with DAC experiments on 19 materials and a
discussion of some possible causes of the existing discrepancies has
recently been published [10] .For example,a phase transition occurs in CdS
at 27 kbar,while the SK prediction is 26.3 kbar .

Work aimed at refining the method for deriving the EOS of a material under
pressure and diminishing the causes of discrepancies discussed in [10] is at
present in progress.\bigskip\ 

Note ( added March 7$^{th}$,1998 ):\medskip\ 

This brief review was published in : AIP Conference Proceedings
Series,Vol.309,p.53 ( 1994 ). Author P.Savic died in May,1994. For a longer
review of the SK theory , and a brief account of the life of P.Savic,the
reader is reffered to the paper by V.Celebonovic: Bull.Astron.Belgrade,{\bf %
151}, 37 \qquad (1995 ). Results obtained within the SK theory on the cold
compression curve of dense matter are presented in: Publ.Astron.Obs.Belgrade,%
{\bf 54}, 203\qquad ( 1996 ). Both papers exist at http://xxx.lanl.gov,in
the astro-ph and cond-mat archives .Recent results obtained by the Mars
Pathfinder on the internal structure of Mars ( published in "Science" in 
December 1997.) are also in agreement with the predictions of this theory.
\bigskip\ 

References

[1] P.Savic,Bull.de la classe des Sci.Math.et Natur.de l'Acad.Serbe des
Sciences et des Arts,{\bf 20}, 107 ( 1961 ).

[2] P.Savic and R.Kasanin,The Behaviour of Materials Under High Pressure
I-IV,Beograd:\ SANU ( 1962/65).

[3] D.Ma,Z.Wang,J.Chen and C.Zhang, J.Phys.,C{\bf 21}, 3585 ( 1988 ).

[4] S.R.de Groot and L.G.Suttorp,Foundations of Electrodynamics,North
Holland Publ.Comp.,Amsterdam ( 1972 ).

[5] P.Savic,Adv.Space.Res.,{\bf 1},131 ( 1981 ).

[6] V.Celebonovic,Earth,Moon and Planets,{\bf 34,} 59 ( 1986 ).

[7] V.Celebonovic,ibid,{\bf 42}, 297 ( 1988 ).

[8] V.Celebonovic,ibid,{\bf 45}, 291 ( 1989 ).

[9] V.Celebonovic,ibid,{\bf 54}, 145 ( 1991 ).

[10] V.Celebonovic,ibid,{\bf 58}, 203 ( 1992 ).{\bf \ }

\end{document}